# A Light Weight Cryptographic Solution for 6LoWPAN Protocol Stack

Sushil Khairnar

Electronics and Telecommunications
Pune Institute of Computer Technology
Pune, India

Gaurav Bansod

Electronics and Telecommunications
Pune Institute of Computer Technology
Pune, India

Vijay Dahiphale

Electronics and Telecommunications
Pune Institute of Computer Technology
Pune, India

*Abstract*—**Lightweight cryptography is an emerging field in the field of research, which endorses algorithms which are best suited for constrained environment. Design metrics like Gate Equivalence (GE), Memory Requirement, Power Consumption, and Throughput play a vital role in the applications like IoT. This paper presents the 6LoWPAN Protocol Stack which is a popular standard of communication for constrained devices. This paper presents an implementation of a lightweight 6LoWPANProtocolstack by using a Light weight Cipher instead of regular heavy encryption cipher AES. The cipher proposed in this paper is specifically suitable for 6LoWPAN architecture as it addresses all the constraints possessed by wireless sensor nodes. The lightweight cipher proposed in the paper needs only 1856 bytes of FLASH and 1272 bytes of RAM memory which is less than any other standard existing lightweight cipher design. The proposed cipher's power consumption is around 25mW which is significantly less as compared to ISO certified lightweight cipher PRESENT which consumes around 38 mW of dynamic power. This paper also discusses thedetailed analysis of cipher against the attacks like Linear Cryptanalysis, Differential Cryptanalysis, Biclique attack and Avalanche attack. The cipher implementation on hardware is around 1051 GE's for 64 bit of block size with 128 bit of key length which is less as compared to existing lightweight cipher design. The proposed cipher LiCi-2 is motivated from LiCi cipher design but outclasses it in every design metric. We believe the design of LiCi-2 is the obvious choice for researchers to implement in constrained environments like IoT.**

*Keywords*—**6LoWPAN, IOT, Light weight Cryptography.**

## I. INTRODUCTION

Cryptography is a science of formulating methods and associated procedures, for transforming data, using algorithm so as to make it unreadable to anyone, except those possessing the key. Light weight Cryptography is a subset of cryptography, which deals with constrained environment, involving less number of GE's, less Computational Power, less Memory Requirement and predominantly less Footprint area. Light weight Cryptography finds its applications in the field of IOT, Sensors, Smart Cards, RFID Tags, Health Care, etc. For these applications minimal GE's are required (GE's < 2200). Since these devices operate in constrained environment, therefore the traditional ciphers like AES are not suitable to do the job. Hence in recent years the demand for Light weight ciphers has augmented. In general a Cipher is primarily divided into 2 parts, Block ciphers and Stream ciphers.

Block ciphers usually operate on a group of bits called Blocks and each block is processed multiple times. A unique key is applied to each block in every round and eventually resulting in a secure cipher text. The main goal of these types of ciphers is Security, whereas Stream ciphers operate on a block of data, one bit at a time. Pseudo random bits are used for encryption and decryption. A popular application of Stream ciphers is A5/1 Cipher, which is used for voice encryption in GSM mobile phone. Block cipher is further classified into SPN (Substitution and Permutation Network) and Feistal structure. SPN is a special type of block cipher which takes a block of plain text along with the key and undergoes several rounds of substitution layers and permutation layers to produce a secure cipher text. SPN processes 64 bit block of data at a time unlike Feistal structure. AES, PRESENT[1], 3 WAY, SHARK, and SQUARE are some of the popular ciphers which uses SPN. However in the Feistal Structure, the plaintext is divided into 2 symmetric halves and each half of plain text undergoes several rounds of substitutions and permutations like circular shifting, X-ORing, etc. RC2, SIMON and SPEC[2] and CLEFIA[3], are some of the popular ciphers using Feistal structure. Feistal structure is further divided classified into 2 types, Generalized Feistal Structure (GFS) and Classical Feistal Structure (CFS). If the plain text is divided into two or more halves, then it is called as Generalized Feistal Structure (GFS) and if Plain text is divided into exactly 2 halves then it is called as Classical Feistal Structure (CFS). CLEFIA[3] and PICCOLO[4]are the ciphers using GFS.

SPN uses 64 bit of permutation, so throughput is less. However it requires more number of GE's. Feistal structure on the other hand, due to use of block shuffling, has higher throughput. It requires less number of GE's compared to other networks. As the block of data is divided, it uses less memory and provides best security.

Cryptographic algorithms can be symmetric and asymmetric. Symmetric algorithm uses a single private key for communication. Same key is shared between the sender and receiver for communication. Advantage of symmetric algorithm is that it requires less number of keys with less key size.

RECTANGLE[5] cipher has strong SBOX and strong cryptanalysis properties. The lightweight symmetric algorithms used are, AES, HIGHT[6], LED[7] PRESENT[1], RC5, TEA. Whereas, Asymmetric key algorithm uses public and private key for communication. Data is encrypted from public key and decrypted by private key. However, the key size is large in asymmetric algorithm which in turn increases the complexity of algorithm and makes it slower as compared to former. Commonly used light weight asymmetric algorithms are, RSA (Ron Rivest, Adi Shamir, Adlemen) and ECC(Elliptic Curve Cryptography).

This paper presents a Classical feistal design Cipher called LiCi-2. LiCi-2 uses 64 bit block length and 128 bit key. The lightweight cipher proposed in the paper needs only 1856 bytes of FLASH and 1272 bytes of RAM memory. It requires only 1051 GEs and the power consumption is 25mW, which makes it suitableto work in constraint environment.This paper gives a brief discussion about the 6LoWPAN Protocol Stack which is a popular standard of communication for constrained devices.This paper also discusses the application of LiCi-2 cipher in 6LoWPAN Protocol stack for its optimum performance. Chapter VII gives comparision of LiCi-2 with the existing light weight Cipher LiCi. LiCi-2 cipher attains good security against the attacks like Linear and Differential attack,Biclique attack and Avalanche attack.

## II. IOT SECURITY AND 6LOWPAN

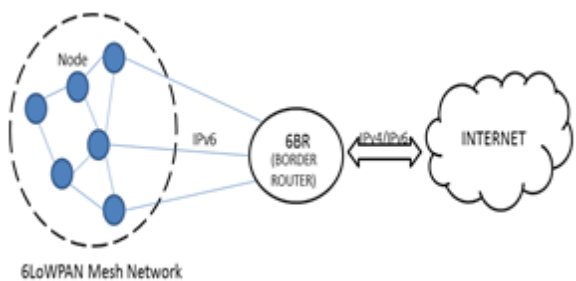

Fig. 1. 6LoWPAN and connection to Internet

In the last few years, IoT has been the hot topic of research and development, as every device is getting smarter day by day.

Smart devices like laptop, phone, TV, refrigerator, AC, charger, and many more are connected to internet. IoT can be defined as a network of uniquely identifiable, accessible, and manageable smart things that are capable of communication, computation and ultimate decision making[8]. The things in IoT can be connected using wireless connections like RFID, Bluetooth, ZigBee, WSN, WLAN, WMAN or Wi-Fi. But for the safe communication between sender and receiver IOT security is essential.

The IOT security relies on certain important parameters which include:

**Confidentiality**: Data can be accessed by only sender or receiver.

**Data Integrity**: The Intruder cannot modify our transmitted data

**Authentication**: The identity of the sender should be unique and should be verified at the receiver end to judge the validity of data.

**Authorization**: Only valid users are able to access the IoT resources. But since the main focus of this paper is towards the security for light weight (constrained) devices, we will be discussing the security for such devices.

IoT can be secured at various layers:

**IEEE 802.15.4** defines Physical and data link layers used for the Low Rate Wireless Personal Area Networks (LRWPAN) using short distance applications with low power and low cost communication networks, particularly for the short range applicationssuch as Wireless Sensors Network (WSN) [9]. IP security (**IPsec**) defines the network layer. In built cryptographic protocols like AES, etc are implemented at this layer and **DTLS**( Datagram Transport Layer Security) can be implemented at the transport layer.DTLS can be used for automatic key management. Data Integrity and authentication can also be provided with the help of DTLS.

6LoWPAN is a low power wireless network wherein each node has its own IPv6 address, using which it can connect directly to the internet using open standards (Fig 1.)

LoWPANs is the term used for low power and lossy networks. LoWPANs have various characteristics like Limited Processing Capability, Low Cost, Small Memory Capacity (few kilobytes of RAM and few kilobytes of flash memory), Low Power (in the order of mW) and Short Range.

IPv6 over Low Power WPAN (**6lowpan**) is an AES standard used for security architecture in constrained environment. 6lowpan integrates IP based infrastructures and WSN and uses header compression mechanisms like IPHC (Internet Protocol Header Compression), NHC (Next Header Compression), UDP Header, GHC (Generic header compression). IPv6 packets can be routed in constrained environment like IEEE 802.15.4. [10] 6LoWPAN protocol uses asynchronous mode of communication. It adopts a mesh topology and uses a routing algorithm which does not bothers about sleeping node, thus it requires an approach such as low-power listening for energy saving purpose. In order to deploy an IoT network, installing a Border Router (BR) is a must. Border Router is a gateway that connects 802.15.4 (WPAN) devices which are on one side of it, and Ethernet or Wi-Fi is on other side of it. 6LoWPAN uses a border router named as 6LBR.

6LoWPAN advantages:

- It uses Open standards including TCP, UDP, HTTP, COAP, MQTT, and websockets.
- End-to-end IP addressable nodes.
- Mesh routing is possible which can be of the form one-to-many or many-to-one.
- Self-healing
- Robust and Scalable
- Multiple PHY support
- Interoperability at the IP level

The basic structure of 6LoWPANProtocol stack is shown in the Fig 2.

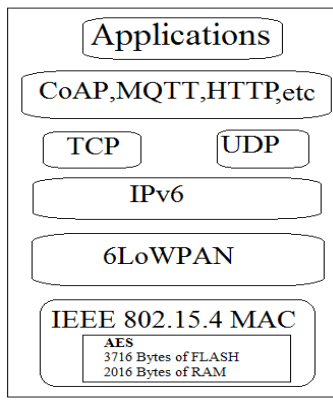

Fig. 2. 6LoWPAN Protocol Stack

Challenges in 6LoWPAN:
- The Header Overhead is large: 802.15.4 has maximum frame overhead of 25 bytes. Link layer security can be high as 21 bytes. This leaves 81 bytes left. Out of which 40 bytes are for IP header and 8 bytes for UDP header. Thus only 33 bytes are remaining for actual data. This is less than desirable.

- IPsec is mandatory with IPv6, considering power constraints. And it is observed that IPsec is computationally expensive.
- 6LoWPAN has limited packet size, which implies that headers for IPv6 and layers above must be compressed whenever possible.
- 6LoWPANs require simple service discovery network protocols to discover, control and maintain services provided by devices.
- 6LoWPAN is susceptible to physical attacks, that is, threats due to physical node destruction relocation and masking.
- 6LoWPAN face the problem of Sleep Deprivation, in which the battery depletes at faster rate and 6LoWPAN devices try to sleep as often as possible in order to conserve it.

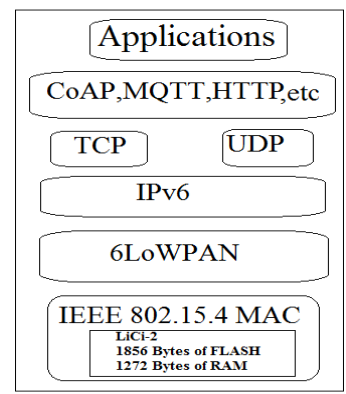

Fig. 3. Light Weight 6LoWPAN Protocol Stack

6LoWPAN uses AES standard used for security, because of which the Protocol Stack of 6LoWPAN is slightly heavier and can be optimized by any of the two techniques discussed below. AES algorithm needs more processing time and memory requirement is more (3716 bytes of FLASH memory and 2016 bytes of RAM).

These memory requirements can be minimized by the cipher proposed by us which requires only 1856 bytes of FLASH memory and 1272 bytes of RAM (Fig. 3). Because of the decrease in memory, the load on the 6LoWPAN protocol stack may minimize to some extent and the processing would be faster.

Other possible solution to increase the processing speed is by using Intellectual Property Core, wherein we can implement the encryption algorithm as IP core on the hardware platform and then transmit the data via the system bus or using any serial communication protocol.

As it is observed that the 6LoWPAN node dissipates its energy at a faster rate because the power consumption is large for encrypting data using AES. The power consumption can be minimized by reducing the number of Gate Equivalents. Hardware implementation of AES requires around 2400 GEs. Higher gate count in AES is mainly because of the hardware complexity of AES components, in particular its S-box.[2]

When resources used are minimal, in most cases it will not use an algorithm providing 128 (or 192 or 256) bits of block size. The AES algorithm uses a block size of 128 bits which is not always optimal. For many small scale embedded applications like for example if a 6LoWPAN communication protocol requires only 64-bit quantities to be encrypted, and demanding 128 bits , as a result of which there is an unnecessary wastage of chip area.

The proposed cipher LiCi-2 addresses all these issues.LiCi-2 uses 64 bit block length and requires only 1051 GEs which can be very efficient in constraint environment.

### III. SPECIFICATIONS OF LiCi-2

- P - 64-bit input plaintext
- C - 64-bit output cipher text
- K - 128-bit Round sub key for round i
- F1 - Function
- $\oplus$ - Bitwise exclusive-OR operation
- <<<x Left cyclic shift by x bits
- >>>x Right cyclic shift by x bits
- RCi Round counter i
- | - Concatenation
- 64 bits Maximum length of plain text
- MSB - Most Significant Bits
- LSB - Least Significant Bits
- ← - Assignment operator

### IV. LiCi-2 CIPHER DESIGN

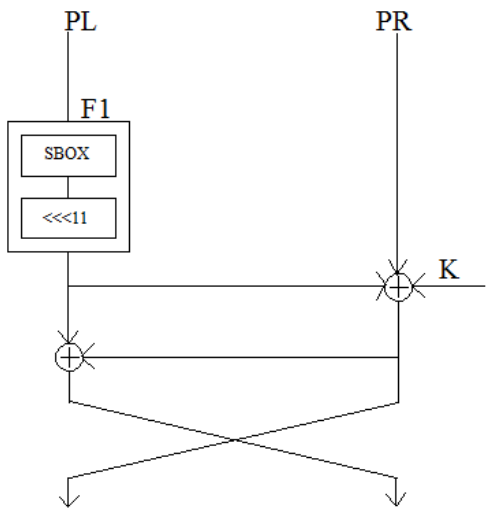

Fig. 4. A Block Cipher LiCi-2

The block cipher LiCi-2 is shown in (Fig. 4) . The 64 bit plain text is divided into 2 equal halves PL and PR, where PL comprises of upper 32 bits of plain text (P63,P62,P61,……P33,P32) and PR comprises of upper 32 bits of plain text (P31,P30,P29,…….P1,P0). The PL is passed through function F1 wherein, PL is first operated with substitution layer and then followed by a left cyclic shift by 11. The result is ex-ored with PR and round key. The resultant of this ex-or operation is again ex-ored with the output of F1. Finally the contents of both the ex-ors are swapped and the same cycle repeats for 25 rounds.

#### A. Design Rationale of LiCi-2

The design of LiCi-2 consists of two XOR operations, one function F1(comprising of 1 S-BOX and a Left Circular Shift operation) and swapping operation . The performance of LiCi-2 is optimum for hardware as well as software platform.

In the design of our cipher, we have not used bit permutation block as it uses additional FLASH memory as well as RAM .The GEs required for Bit permutation block are more which makes the design a bit bulky. Though bit permutations theoretically don't require any GE'S but practically it results in gate overhead because of storing intermediate values in the registers.

S- BOX is the non-linear block of our cipher design. We have chosen an S-BOX in such a way that it is Light weight and also ensures optimum security.

Left Circular shift by 11 bits was chosen so as to get maximum number of active S- Boxes which can easily resist linear and differential attacks. Two XOR operations are used in place of bit permutation blocks which will eventually reduce the number of GEs.

Power consumption of this cipher when implemented on hardware is minimal (25mW).It is because of limiting number of rounds to 25. Moreover due to simple design, very less number of local and global variables are used which also reduces power consumption.

 It requires only 1856 bytes of RAM because of which software performance is excellent as it can be observed from the Execution time(150.1 usec). It also generates significantly higher throughput due to processing of 32 bit block size at a given time.

Because of all these design choices, LiCi-2 is the simplest design of a light weight cipher till date without any compromise on hardware and software performance and on security issues.

### B. Encryption algorithm

1. Plain text is divided into 2 equal halves where,
   $P_i{}^R$= MSB 32 bits (P63,P62,P61,.......P33,P32)
   $P_i{}^L$= LSB 32 bits (P31,P30,P29,.......P1,P0)
   $P \leftarrow P_i{}^R| P_i{}^L$
2. Apply F1 to $P_i{}^L$
   $P_{i1} \leftarrow$ SBOX ($P_i{}^L$)
   $P_{i2} \leftarrow P_{i1} <<< 11$
3. EX-OR $P_{i2}$ with $P_i{}^R$ and Round key $K_i$
   $P_{i3} \leftarrow P_{i2} \oplus P_i{}^R \oplus K_i$
4. EX-OR $P_{i3}$ with $P_{i2}$
   $P_{i4} \leftarrow P_{i3} \oplus P_{i2}$
5. Block shuffle $P_{i4}$ and $P_{i3}$ by 32 bits
   $P_{i+1}{}^L \leftarrow P_{i3}$
   $P_{i+1}{}^R \leftarrow P_{i4}$
6. After 25 rounds, we will get cipher text as:
   $C \leftarrow P_{25}{}^L| P_{25}{}^R$

### C. SBOX

S-Box is a crucial block in the design of cipher, as it is the only non-linear block in the cipher. Strong S-Box makes the cipher secure and helps to attain good security level in a minimum number of rounds. S-BOX used in LiCi-2 cipher design is a 8 − 4 bit S-BOX. The SBOX operates on MSB 32 bits of plain text. Table I shows the hexadecimal values for substitution layer.

TABLE I. S-BOX OF LiCi-2 CIPHER

| X | 0 | 1 | 2 | 3 | 4 | 5 | 6 | 7 | 8 | 9 | A | B | C | D | E | F |
|---|---|---|---|---|---|---|---|---|---|---|---|---|---|---|---|---|
| S(x) | 3 | F | E | 1 | 0 | A | 5 | 8 | C | 4 | B | 2 | 9 | 7 | 6 | D |

### D. Key scheduling for 128 Bit Key Length

LiCi-2 Cipher uses key scheduling algorithm which is inspired from PRESENT cipher[1]. PRESENT cipher key scheduling algorithm is robust and till date no attacks are reported. LiCi-2 cipher key scheduling algorithm produces 25 sub keys, used with ex-or operation at right hand side as shown in figure. The key scheduling algorithm for 128 bit key-length is given as:

1. key<<< 13.
2. $[k_3 k_2 k_1 k_0] \leftarrow s [k_3 k_2 k_1 k_0]$
3. $[k_7 k_6 k_5 k_4] \leftarrow s [k_7 k_6 k_5 k_4]$
4. $[k_{63} k_{62} k_{61} k_{60} k_{59}] \leftarrow [k_{63} k_{62} k_{61} k_{60} k_{59}] \oplus RC_i$

## V. HARDWARE AND SOFTWARE PERFORMANCE OF LiCi-2 CIPHER

The performance of LiCi-2 Cipher is computed by taking various hardware and software performance parameters into consideration like Memory Requirement (Flash and RAM), Power Requirement, GE (Gate Equivalent), Execution time and Throughput. The less number of GEs results in a smaller footprint area.

### A. Memory Requirements

For software performance analysis of LiCi-2 cipher we have used 32-bit ARM7 LPC2129 processor. The memory used by our Cipher is 1856 bytes (Flash) and 1272 bytes (RAM). (Fig. 5) represents the memory comparison of the existing lightweight ciphers with the LiCi-2 cipher.

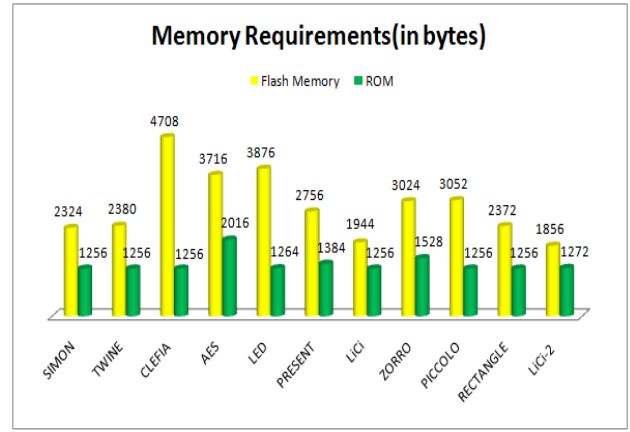

Fig. 5. Memory Requirements Comparison of LiCi-2 with existing Light Weight Ciphers

### B. Power Consumption

We have calculated the power consumption by using Verilog in Xilinx ISE Design Suite 13.2. Power is calculated with on Vertex 6 family and the device used is XC6SLX9 with package CSG324. Graph represents the dynamic power consumption of standard ciphers and its comparison with LiCi-2 cipher. LiCi-2 cipher consumes only 25mW of dynamic power which is lesser compared to other lightweight ciphers.(Fig. 6.) PRESENT consumes 38mW of power, LiCi consumes 30mW of power. RECTANGLE consumes 31mW of power and LED consumes 100mW of power. [11].

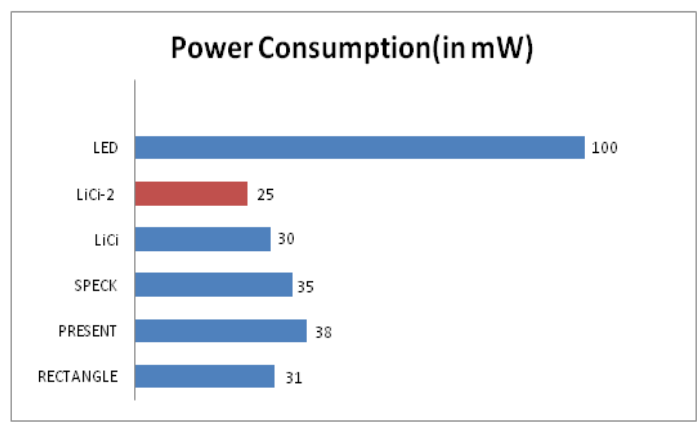

Fig. 6. Power Consumption Comparison of LiCi-2 with existing Light Weight Ciphers on Xilinx Platform.

## C. Execution Time and Throughput

We have used a 32-bit ARM7 LPC2129 processor on Keil µ5 platform for computation of Execution time and Throughput. The execution time for LiCi-2 cipher is 150.1 µsec, which very less compared to other existing ciphers. The throughput computed is 426.38 Kbps. Table II. gives the comparison of Execution time and Throughput with other existing ciphers.

TABLE II.    COMPARISON OF EXECUTION TIME AND THROUGHPUT

| Ciphers | Execution Time (in µsec) | Throughput(in Kbps) |
|---|---|---|
| LED | 7092.86 | 9 |
| PRESENT | 2468.65 | 24.16 |
| SIMON | 105.67 | 605 |
| SPECK | 49.02 | 1305 |
| CLEFIA | 1048.01 | 122 |
| PICCOLO | 227.68 | 281 |
| TWINE | 592.87 | 108 |
| LiCi | 209.87 | 305 |
| **LiCi-2** | **150.1** | **426.38** |

## D. Gate Equivalent

Gate Equivalent gives the total foot print size required for the cipher. We have computed GE's with ARM standard cell library for the IBM 8RF (0.13 µm). The foot print area required for some basic gates is as shown in TABLE III:

TABLE III.    FOOTPRINT AREA OF BASIC GATES

| GATES | AREA(in µm) |
|---|---|
| AND | 1.25 |
| OR | 1.25 |
| XOR | 2 |
| NOT | 0.75 |
| 2:1 MUX | 2.25 |
| D FLIP FLOP | 4.25 |

- GEs Calculation for Data Path:

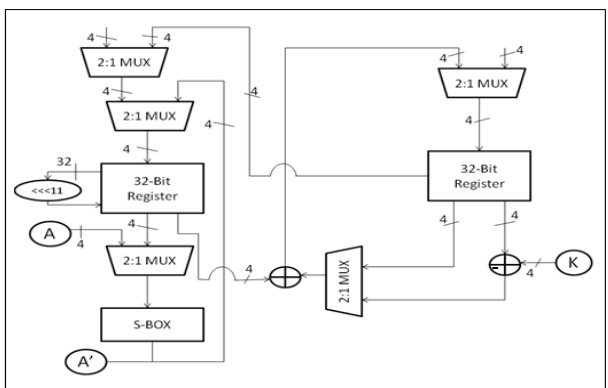

Fig.7. Data Path for LiCi-2 Cipher for 128 bit Key Scheduling

TABLE IV.    GEs REQUIRED FOR DATA PATH

| DATA LAYER | GE'S REQUIRED |
|---|---|
| 64-Bit Register | 64*4.25=272 |
| 2 XOR | 2*4*2=16 |
| 5 MUX | 5*4*2.25=45 |
| 1 SBOX | 24 |
| 1 Shift | 0 |
| **TOTAL** | 357 |

- GEs Calculation for Key Path:

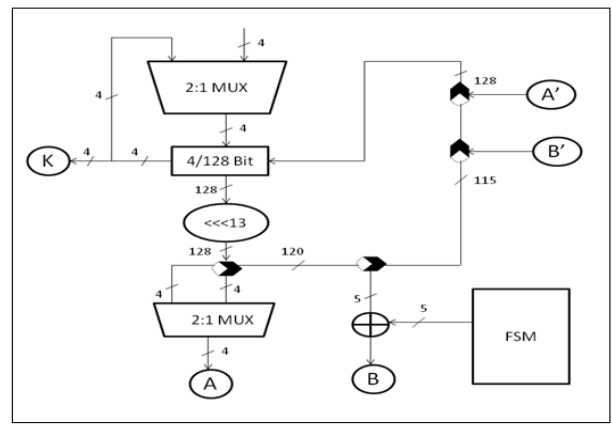

Fig.8.  Key Path for LiCi-2 Cipher for 128 bit Key Scheduling

TABLE V.    GEs REQUIRED FOR KEY PATH

| KEY LAYER | GE'S REQUIRED |
|---|---|
| 128-Bit Register | 128*4.25=544 |
| 1- 5 Bit XOR | 2*5=10 |
| 2 MUX | 2*4*2.25=18 |
| FSM | 122 |
| **TOTAL** | 694 |

Thus, Total number of GE's (calculated from Fig.7 and Fig. 8) required are:

      357 => for Data path (TABLE IV.)
 + 694=> for Key path (TABLE V.)
 = **1051.**

Fig. 9 shows the GEs comparison of LiCi-2 cipher with existing ciphers in terms of GE's.

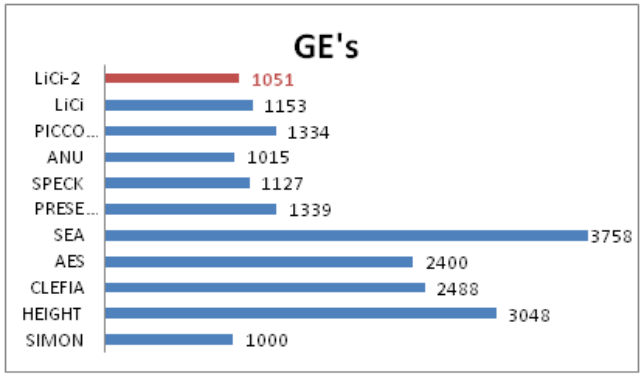

Fig.9. GEs Comparison of LiCi-2 with existing Ciphers

## VI. Sequrity Analysis of LiCi-2

Cryptology is classified into two parts, Cryptography and Cryptanalysis. We have been discussing Cryptography so far, which is mostly used for hiding data. However, Cryptanalysis is a process of analyzing the Cryptosystem and breaking it. Thus using cryptanalysis, the hackers or attackers can retrieve the original data by breaking the key. Thus our cipher should be robust to the attacks. This paper focuses on the attacks like - Linear Cryptanalysis, Differential Cryptanalysis, Biclique attack and Avalanche attack. In order to withstand the basic attacks, the SBOX should be highly non-linear.

### A. Linear Cryptanalysis

Linear cryptanalysis is considered as one of the basic attacks and our design should always be robust against this attack, as it involves the number of active s-boxes used by the cipher design. Active S-box can be defined as the S-box which gives non-zero output for the non-zero input. This attack is also known as known plaintext attack and cipher should fight against such attack. It uses the high probability occurrences of linear expression containing plaintext bits, cipher text bits and sub key bits[11]. The S-box is examined by generating a Linear Approximation Table (LAT) . If pl is the linear probability, then its bias is given as |pl-1/2|, bias ($\varepsilon$). For this cipher bias is $2^{-2}$. Matsui's Piling-up is used to calculate probability bias for 25 rounds [13]. Table VI represents the minimum number of active S-boxes from linear trails.

TABLE VI.     Minimum number of active s- boxes for linear trails

| Number | Number of Active Sboxes |
|--------|-------------------------|
| 1 | 0 |
| 2 | 1 |
| 3 | 3 |
| 4 | 7 |
| 5 | 13 |

**Theorem 1:** For 20 rounds LiCi-2 cipher has 52 active S-boxes and total bias ($\varepsilon_1$) is $2^{-53}$.

**Proof:**
From the table, it can be interpreted that for 5 rounds, the total number of active s-boxes are 13. Thus on approximating (taking the multiplier factor as 4), it can be found that for 20 rounds, the total number of active s-boxes required are 13*4=52. By Matsui's Pilling up Lemma principle for 4 rounds of LiCi-2 cipher, total bias ($\varepsilon_0$) is given as:

$\varepsilon_{0} = 2^{(\text{number of active s-boxes-1})} * (2^{-2})^{(\text{number of active sboxes})}$

$\varepsilon_{0} = 2^{(13-1)} * (2^{-2})^{(13)} = 2^{-14}$

By applying same lemma for 20 rounds, the total bias ($\varepsilon_1$) is given as

$\varepsilon_{1} = 2^{(\text{multiplier-1})} * (\varepsilon_0)^{(\text{multiplier})}$

$\varepsilon_{1} = 2^{(4-1)} * (\varepsilon_0)^{(4)} = 2^{-53}$

The complexity of linear attack is given as
$N_L = 1/(\varepsilon_1)^2$

Complexity defines the required number of known plain text or cipher text. Thus, for 20 rounds the required number of plain text or cipher text is:

$N_L = 1/(\varepsilon_1)^2 = 1/(2^{-53})^2 = 2^{106}$

It can be observed that $2^{106}$ is much greater than $2^{64}$, Hence LiCi-2 is robust against the linear attack and it is secure for 20 rounds. If it is secure for 20 rounds, it has to be secure for any number of rounds greater than 20. Thus LiCi-2 can provide optimum security in just 25 rounds.

### B. Differential Cryptanalysis

Differential attack is also a basic and a significant attack. To mount the differential attack for a specific number of rounds in an encryption system, pairs of high probability input and output occurrences are used to recover the round keys[12]. We make analysis of Differential attack by forming a Difference Distribution TABLE (DDT). Differential trails are formed by considering high probability input and output for each round. Differential probability for the LiCi-2 cipher S-box is 4/16 =1/4 = $2^{-2}$. Table VII represents the minimum number of active S-boxes from differential trails.

TABLE VII.     Minimum number of active s- boxes for Differential trails

| Number of rounds | Number of Active Sboxes |
|------------------|-------------------------|
| 1 | 0 |
| 2 | 1 |
| 3 | 3 |
| 4 | 6 |
| 5 | 10 |

For 5 rounds, there are 10 minimum active S-boxes. So by approximation, for 20 rounds, there will be a minimum of 40 active S-boxes.

Total differential probability (pd) is given as
pd= $(2^{-2})^{(\text{no. of active s-boxes})}$
pd= $(2^{-2})^{40} = 2^{-80}$.

The complexity of the differential attack can be found out by calculating the required number of chosen plaintext or cipher text and can be given as,

Nd= C/Pd

Where C = 1 and Pd= $2^{-80}$, so the required number of chosen plaintext or cipher text are
Nd=1/($2^{-80}$)= $2^{80}$.

It can be observed that $2^{80}$ is much greater than $2^{64}$, Hence LiCi-2 is robust against the differential attack and it is secure for 20 rounds and thus for 25 rounds.

### C.  Biclique Attack

In this paper, we have applied Biclique attack to LiCi-2 Cipher with a 128 bit key, from round 22 to 25 and it is a 4 dimensionalBiclique Attack. This type of attack decides the data complexity of the cipher. For the forward computation,we have used the red key for rounds 23, 24 and 25.For the respective key positions we got the values as:
K[6]=91
K[5]=90
K[4]=89
K[3]=88
And for the the backward computation we have used the blue key for rounds 24, 23, and 22. For the respective key positions we got the values as:
K[29]=101
K[28]=100
K[27]=99
K[26]=98
Using these results we computed the data complexity of the biclique attack as 2^50.

### D.  Avalanche Attack

Significant change in output with change in single bit at input plaintext, this result in avalanche effect [13]. We have designed LiCi-2 cipher for 25 rounds with 64 bit block length and 128 bit of key length. So when we apply data and a key to a cipher, it should be observed that there must be at least 50% change in the number of bits in cipher text for a single bit change in data or a key. If there is 50% change (32 bits in our case) in the number of bits then this phenomenon is called as Strict Avalanche. The number of bits changed depends on the design of the cipher that is on the number of XOR operations used or on the circular shifts and even on the bit permutations. Our cipher successfully overcomes the avalanche attack and is robust against this attack.

VII.        COMPARISON BETWEEN LiCi AND LiCi-2

Fig. 10 Represents the comparison between LiCi[14] and LiCi-2. Here we have considered the parameters of LiCi as 100 percentile. So based on the observations, we have concluded that the FLASH Memory of LiCi-2 is about 5% lesser than that of LiCi. Similarly GEs required are around 9.7% lesser, Power Consumption is 20% lesser,

Execution time is 37.18% lesser and Throughput is around 39.79% greater in LiCi-2 when compared to LiCi. The detailed comparison is given in Table VIII.

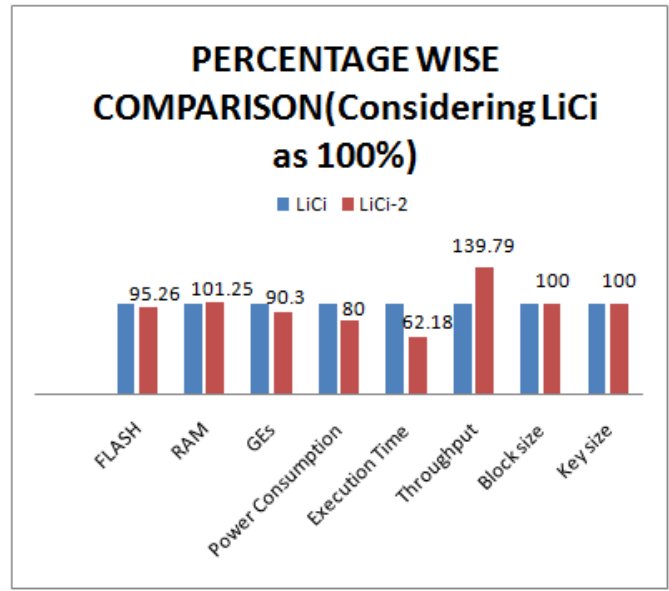

Fig.10. Statistical Comparison of LiCi and LiCi-2

TABLE VIII.      COMPARISON OF LiCi AND LiCi-2

| Parameters | LiCi | **LiCi-2** |
|---|---|---|
| Memory (in bytes):<br>FLASH<br>RAM | 1944<br>1256 | **1856**<br>**1272** |
| GEs | 1153 | **1051** |
| Power Consumption (in mW) | 30 | **25** |
| Execution Time (in μsec ) | 206.87 | **150.1** |
| Throughput (in Kbps) | 305 | **426.38** |
| Block size (in bytes): | 64 | **64** |
| Key size (in bytes): | 128 | **128** |

## VIII. CONCLUSION

This paper presents the role of Light weight Cryptography in architectures like 6LoWPAN and IoT. It presents a Classical Feistal structure design named as LiCi-2 which operates on 64 bit block length and 128 bit key length. This paper gives a brief discussion about 6LoWPAN, its advantages, drawbacks and how can it be optimized with the help of LiCi-2 cipher. LiCi-2 Cipher uses very less memory(1856 bytes of Flash and 1272 bytes of ROM) and it requires a very minimal power (25mW). LiCi-2 is motivated from LiCi cipher but performs efficiently on software and hardware platforms and has better design metrics than LiCi cipher. LiCi-2 also shows good resistance against the advanced attack like Biclique and also results in good avalanche property. LiCi-2 has an innovative architecture which helps to generate a maximum number of active S- Boxes in the minimum number of rounds though the encryption design structure is simple in nature. LiCi-2 gives excellent hardware as well as software performance, which makes it the best choice for using in any small scale embedded applications.LiCi-2 cipher requires only 1051 GEs which makes it an ideal choice for various IOT applications. The Test Vectors of LiCi-2 Cipher are shown in TABLE IX. The cipher text is calculated for 2 sets of inputs.

TABLE IX.    Test Vectors of LiCi-2 with 128 bit key.

| Plain Text | Key | Cipher Text |
|---|---|---|
| 12345678 90abcdef | 12345678 90abcdef 12345678 90abcdef | 1339607b 88df737a |
| FFFFFFFF FFFFFFFF | 00000000 00000000 00000000 00000000 | c7ac349c ecb57df3 |